\documentclass[aps,prd,twocolumn,superscriptaddress,showpacs]{revtex4}

\usepackage{epsfig}
\usepackage{graphicx}
\usepackage{dcolumn}
\usepackage{bm}
\usepackage{setspace}

\begin{document}


\renewcommand{\thetable}{\Roman{table}}


\title{
Observation of the $\Omega^-_b$ Baryon and
Measurement of the Properties of the $\Xi^-_b$ and $\Omega_b^-$ Baryons}
\affiliation{Institute of Physics, Academia Sinica, Taipei, Taiwan 11529, Republic of China} 
\affiliation{Argonne National Laboratory, Argonne, Illinois 60439} 
\affiliation{University of Athens, 157 71 Athens, Greece} 
\affiliation{Institut de Fisica d'Altes Energies, Universitat Autonoma de Barcelona, E-08193, Bellaterra (Barcelona), Spain} 
\affiliation{Baylor University, Waco, Texas  76798} 
\affiliation{Istituto Nazionale di Fisica Nucleare Bologna, $^y$University of Bologna, I-40127 Bologna, Italy} 
\affiliation{Brandeis University, Waltham, Massachusetts 02254} 
\affiliation{University of California, Davis, Davis, California  95616} 
\affiliation{University of California, Los Angeles, Los Angeles, California  90024} 
\affiliation{University of California, San Diego, La Jolla, California  92093} 
\affiliation{University of California, Santa Barbara, Santa Barbara, California 93106} 
\affiliation{Instituto de Fisica de Cantabria, CSIC-University of Cantabria, 39005 Santander, Spain} 
\affiliation{Carnegie Mellon University, Pittsburgh, PA  15213} 
\affiliation{Enrico Fermi Institute, University of Chicago, Chicago, Illinois 60637}
\affiliation{Comenius University, 842 48 Bratislava, Slovakia; Institute of Experimental Physics, 040 01 Kosice, Slovakia} 
\affiliation{Joint Institute for Nuclear Research, RU-141980 Dubna, Russia} 
\affiliation{Duke University, Durham, North Carolina  27708} 
\affiliation{Fermi National Accelerator Laboratory, Batavia, Illinois 60510} 
\affiliation{University of Florida, Gainesville, Florida  32611} 
\affiliation{Laboratori Nazionali di Frascati, Istituto Nazionale di Fisica Nucleare, I-00044 Frascati, Italy} 
\affiliation{University of Geneva, CH-1211 Geneva 4, Switzerland} 
\affiliation{Glasgow University, Glasgow G12 8QQ, United Kingdom} 
\affiliation{Harvard University, Cambridge, Massachusetts 02138} 
\affiliation{Division of High Energy Physics, Department of Physics, University of Helsinki and Helsinki Institute of Physics, FIN-00014, Helsinki, Finland} 
\affiliation{University of Illinois, Urbana, Illinois 61801} 
\affiliation{The Johns Hopkins University, Baltimore, Maryland 21218} 
\affiliation{Institut f\"{u}r Experimentelle Kernphysik, Universit\"{a}t Karlsruhe, 76128 Karlsruhe, Germany} 
\affiliation{Center for High Energy Physics: Kyungpook National University, Daegu 702-701, Korea; Seoul National University, Seoul 151-742, Korea; Sungkyunkwan University, Suwon 440-746, Korea; Korea Institute of Science and Technology Information, Daejeon, 305-806, Korea; Chonnam National University, Gwangju, 500-757, Korea} 
\affiliation{Ernest Orlando Lawrence Berkeley National Laboratory, Berkeley, California 94720} 
\affiliation{University of Liverpool, Liverpool L69 7ZE, United Kingdom} 
\affiliation{University College London, London WC1E 6BT, United Kingdom} 
\affiliation{Centro de Investigaciones Energeticas Medioambientales y Tecnologicas, E-28040 Madrid, Spain} 
\affiliation{Massachusetts Institute of Technology, Cambridge, Massachusetts  02139} 
\affiliation{Institute of Particle Physics: McGill University, Montr\'{e}al, Qu\'{e}bec, Canada H3A~2T8; Simon Fraser University, Burnaby, British Columbia, Canada V5A~1S6; University of Toronto, Toronto, Ontario, Canada M5S~1A7; and TRIUMF, Vancouver, British Columbia, Canada V6T~2A3} 
\affiliation{University of Michigan, Ann Arbor, Michigan 48109} 
\affiliation{Michigan State University, East Lansing, Michigan  48824}
\affiliation{Institution for Theoretical and Experimental Physics, ITEP, Moscow 117259, Russia} 
\affiliation{University of New Mexico, Albuquerque, New Mexico 87131} 
\affiliation{Northwestern University, Evanston, Illinois  60208} 
\affiliation{The Ohio State University, Columbus, Ohio  43210} 
\affiliation{Okayama University, Okayama 700-8530, Japan} 
\affiliation{Osaka City University, Osaka 588, Japan} 
\affiliation{University of Oxford, Oxford OX1 3RH, United Kingdom} 
\affiliation{Istituto Nazionale di Fisica Nucleare, Sezione di Padova-Trento, $^z$University of Padova, I-35131 Padova, Italy} 
\affiliation{LPNHE, Universite Pierre et Marie Curie/IN2P3-CNRS, UMR7585, Paris, F-75252 France} 
\affiliation{University of Pennsylvania, Philadelphia, Pennsylvania 19104}
\affiliation{Istituto Nazionale di Fisica Nucleare Pisa, $^{aa}$University of Pisa, $^{bb}$University of Siena and $^{cc}$Scuola Normale Superiore, I-56127 Pisa, Italy} 
\affiliation{University of Pittsburgh, Pittsburgh, Pennsylvania 15260} 
\affiliation{Purdue University, West Lafayette, Indiana 47907} 
\affiliation{University of Rochester, Rochester, New York 14627} 
\affiliation{The Rockefeller University, New York, New York 10021} 
\affiliation{Istituto Nazionale di Fisica Nucleare, Sezione di Roma 1, $^{dd}$Sapienza Universit\`{a} di Roma, I-00185 Roma, Italy} 

\affiliation{Rutgers University, Piscataway, New Jersey 08855} 
\affiliation{Texas A\&M University, College Station, Texas 77843} 
\affiliation{Istituto Nazionale di Fisica Nucleare Trieste/Udine, I-34100 Trieste, $^{ee}$University of Trieste/Udine, I-33100 Udine, Italy} 
\affiliation{University of Tsukuba, Tsukuba, Ibaraki 305, Japan} 
\affiliation{Tufts University, Medford, Massachusetts 02155} 
\affiliation{Waseda University, Tokyo 169, Japan} 
\affiliation{Wayne State University, Detroit, Michigan  48201} 
\affiliation{University of Wisconsin, Madison, Wisconsin 53706} 
\affiliation{Yale University, New Haven, Connecticut 06520} 
\author{T.~Aaltonen}
\affiliation{Division of High Energy Physics, Department of Physics, University of Helsinki and Helsinki Institute of Physics, FIN-00014, Helsinki, Finland}
\author{J.~Adelman}
\affiliation{Enrico Fermi Institute, University of Chicago, Chicago, Illinois 60637}
\author{T.~Akimoto}
\affiliation{University of Tsukuba, Tsukuba, Ibaraki 305, Japan}
\author{B.~\'{A}lvarez~Gonz\'{a}lez$^t$}
\affiliation{Instituto de Fisica de Cantabria, CSIC-University of Cantabria, 39005 Santander, Spain}
\author{S.~Amerio$^z$}
\affiliation{Istituto Nazionale di Fisica Nucleare, Sezione di Padova-Trento, $^z$University of Padova, I-35131 Padova, Italy} 

\author{D.~Amidei}
\affiliation{University of Michigan, Ann Arbor, Michigan 48109}
\author{A.~Anastassov}
\affiliation{Northwestern University, Evanston, Illinois  60208}
\author{A.~Annovi}
\affiliation{Laboratori Nazionali di Frascati, Istituto Nazionale di Fisica Nucleare, I-00044 Frascati, Italy}
\author{J.~Antos}
\affiliation{Comenius University, 842 48 Bratislava, Slovakia; Institute of Experimental Physics, 040 01 Kosice, Slovakia}
\author{G.~Apollinari}
\affiliation{Fermi National Accelerator Laboratory, Batavia, Illinois 60510}
% ...

\collaboration{CDF Collaboration\footnote{With visitors from $^a$University of Massachusetts Amherst, Amherst, Massachusetts 01003,
$^b$Universiteit Antwerpen, B-2610 Antwerp, Belgium, 
$^c$University of Bristol, Bristol BS8 1TL, United Kingdom,
$^d$Chinese Academy of Sciences, Beijing 100864, China, 
$^e$Istituto Nazionale di Fisica Nucleare, Sezione di Cagliari, 09042 Monserrato (Cagliari), Italy,
$^f$University of California Irvine, Irvine, CA  92697, 
$^g$University of California Santa Cruz, Santa Cruz, CA  95064, 
$^h$Cornell University, Ithaca, NY  14853, 
$^i$University of Cyprus, Nicosia CY-1678, Cyprus, 
$^j$University College Dublin, Dublin 4, Ireland,
$^k$University of Edinburgh, Edinburgh EH9 3JZ, United Kingdom, 
$^l$University of Fukui, Fukui City, Fukui Prefecture, Japan 910-0017
$^m$Kinki University, Higashi-Osaka City, Japan 577-8502
$^n$Universidad Iberoamericana, Mexico D.F., Mexico,
$^o$University of Iowa, Iowa City, IA  52242,
$^p$Queen Mary, University of London, London, E1 4NS, England,
$^q$University of Manchester, Manchester M13 9PL, England, 
$^r$Nagasaki Institute of Applied Science, Nagasaki, Japan, 
$^s$University of Notre Dame, Notre Dame, IN 46556,
$^t$University de Oviedo, E-33007 Oviedo, Spain, 
$^u$Texas Tech University, Lubbock, TX  79609, 
$^v$IFIC(CSIC-Universitat de Valencia), 46071 Valencia, Spain,
$^w$University of Virginia, Charlottesville, VA  22904,
$^x$Bergische Universit\"at Wuppertal, 42097 Wuppertal, Germany,
$^{ff}$On leave from J.~Stefan Institute, Ljubljana, Slovenia, 
}}
\noaffiliation


\begin{abstract}
We report the observation of 
the  bottom, doubly-strange baryon
$\Omega^-_b$ through the decay chain 
$\Omega^-_b \rightarrow J/\psi \, \Omega^-$,
where $J/\psi \rightarrow \mu^+\, \mu^-$, 
$\Omega^- \rightarrow \Lambda \, K^-$,
and $\Lambda \rightarrow p \, \pi^-$,
using 4.2 fb$^{-1}$ of 
data from $p\bar p$ collisions at $\sqrt{s}=1.96$ TeV, and
recorded with the Collider Detector at Fermilab.
A signal is observed whose probability of arising 
from a background fluctuation is $4.0 \times 10^{-8}$, or 
5.5 Gaussian standard deviations.
The $\Omega^-_b$ mass is measured to be 
$6054.4\pm 6.8 (\textrm{stat.})\pm 0.9 (\textrm{syst.})$ MeV/$c^2$.
The lifetime of the $\Omega^-_b$ baryon is measured to be 
$1.13^{+0.53}_{-0.40}(\textrm{stat.})\pm0.02(\textrm{syst.})$ ps.
In addition, for the $\Xi^-_b$ baryon we measure a mass of
$5790.9\pm2.6(\textrm{stat.})\pm0.8(\textrm{syst.})$ MeV/$c^2$
and a lifetime of
$1.56^{+0.27}_{-0.25}(\textrm{stat.})\pm0.02(\textrm{syst.})$ ps.
Under the assumption that the $\Xi_b^-$ and $\Omega_b^-$ are produced with
similar kinematic distributions to the $\Lambda^0_b$ baryon, we find
$\frac{\sigma(\Xi_b^-){\cal B}(\Xi_b^- \rightarrow J/\psi \, \Xi^-)}
      {\sigma(\Lambda^0_b){\cal B}(\Lambda^0_b \rightarrow J/\psi \, \Lambda)} =
0.167^{+0.037}_{-0.025}(\textrm{stat.})\pm0.012(\textrm{syst.})$ and
$\frac{\sigma(\Omega_b^-){\cal B}(\Omega_b^- \rightarrow J/\psi \, \Omega^-)}
      {\sigma(\Lambda^0_b){\cal B}(\Lambda^0_b \rightarrow J/\psi \, \Lambda)} =
0.045^{+0.017}_{-0.012}(\textrm{stat.})\pm0.004(\textrm{syst.})$
for baryons produced with transverse momentum in the range of
 $6 \, - \,  20$ GeV/$c$.
\end{abstract}
\maketitle

\section{Introduction}
Since its inception, the quark model has had great success in describing
the spectroscopy of hadrons.  
In particular, this has been the case for the $D$ and 
$B$ mesons, where all of the 
ground states have been observed \cite{PDG}.
The spectroscopy of $c$-baryons also agrees well with the quark model, 
and a  rich
spectrum of baryons containing $b$ quarks is predicted \cite{Jenkins}.
Until recently, direct observation of $b$-baryons 
has been limited to a single state, the
$\Lambda^0_b$ (quark content $|udb \, \rangle$) \cite {PDG}.  
The accumulation of large data sets from the Tevatron has changed
this situation, and made possible the observation of the
$\Xi^-_b$ ($|dsb \, \rangle$) \cite{D0_Xi_b, CDF_Xi_b}
and the $\Sigma^{(*)}_b$ states ($|uub \, \rangle, |ddb \, \rangle$) 
\cite{CDF_Sigma_b}.

In this paper, we report the observation of
an additional heavy baryon 
and the measurement of its mass, lifetime, and relative production
rate compared to the $\Lambda^0_b$ production.
The decay properties of this state are consistent with the weak decay
of a $b$-baryon.  We interpret our result as the observation
of the $\Omega_b^-$ baryon ($|ssb \, \rangle$). 
Observation of this baryon has been previously reported 
by the D0 Collaobration \cite{D0_Omega_b}.
However, the analysis presented here measures a mass of the $\Omega_b^-$
to be significantly lower than Ref. \cite{D0_Omega_b}.
 
This $\Omega_b^-$ observation is made in
$p\overline{p}$ collisions 
at a center of mass energy of 1.96 TeV using the
Collider Detector at Fermilab (CDF II), through the decay chain 
$\Omega^-_{b} \rightarrow$ $J/\psi \, \Omega^-$, where
$J/\psi \rightarrow \mu^+\mu^-$, $\Omega^- \rightarrow \Lambda \, K^-$,
and $\Lambda \rightarrow p \, \pi^-$.
Charge conjugate modes are included implicitly.
Mass, lifetime, and production rate measurements are also
reported for the $\Xi_b^-$, through the  similar decay chain
$\Xi^-_{b} \rightarrow$ $J/\psi \, \Xi^-$, where
$J/\psi \rightarrow \mu^+\mu^-$, $\Xi^- \rightarrow \Lambda \, \pi^-$,
and $\Lambda \rightarrow p \, \pi^-$.
The production rates of both the $\Xi^-_b$ and $\Omega_b^-$ are measured
with respect to the $\Lambda^0_b$, which is observed through the decay chain
$\Lambda^0_{b} \rightarrow$ $J/\psi \, \Lambda$, where
$J/\psi \rightarrow \mu^+\mu^-$, and $\Lambda \rightarrow p \, \pi^-$.
These measurements are based on 
a data sample corresponding to an integrated luminosity of
4.2 fb$^{-1}$.  

The strategy of the analysis presented here 
is to demonstrate the reconstruction
and property measurements of the $\Xi^-_b$ and $\Omega^-_b$ as 
natural extensions of measurements that can be made on 
better known $b-$hadron states obtained in the same data.
All measurements made here are performed on the 
$B^0 \rightarrow J/\psi \, K^{*}(892)^{0}$, $K^{*}(892)^{0}\rightarrow K^+\pi^-$ 
final state, to provide a large sample
for comparison to other measurements.  
The decay mode $B^0 \rightarrow J/\psi \, K^{0}_s$, 
$K^{0}_s \rightarrow \pi^+ \, \pi^-$ is a second reference
process.  The $K^{0}_s$ is reconstructed  from tracks that
are significantly displaced from the collision, similar to the
final state tracks of the $\Xi^-_b$ and $\Omega^-_b$.
Although its properties are less well measured than those of the
$B^0$, the $\Lambda^0_b \rightarrow J/\psi \, \Lambda$ contributes 
another cross
check of this analysis,
since it is a previously measured state that contains a 
$\Lambda$ in its decay chain.  
The $\Lambda^0_b$ also provides the best state for comparison
of relative production rates, since it is the largest sample
of reconstructed $b$-baryons. 

We begin with a brief description of the detector and its simulation
in Sec.\ \ref{sect:Detector}.
In Sec.\ \ref{sect:Reconstruction}, the reconstruction
of $J/\psi$, neutral $K$, hyperons, and $b-$hadrons is described.
Sec.\ \ref{sect:signif} discusses the extraction and significance of 
the $\Omega^-_b$ signal.  In Sec.\ \ref{sect:Properties}, we present 
measurements of the properties of the $\Xi^-_b$ and $\Omega^-_b$, which
include particle masses, lifetimes, and production rates. 
We conclude in Sec.\ \ref{sect:Conclusions}.

\section{Detector Description and Simulation\label{sect:Detector}}

The CDF II detector has been described in detail elsewhere 
\cite{CDF_detector}.  
This analysis primarily relies upon the tracking and 
muon identification systems.  
The tracking system consists of 
four different detector subsystems that operate inside a 1.4 T solenoid.
The first of these is a single layer of silicon detectors (L00) 
at a radius of $1.35 - 1.6$ cm from the axis of the solenoid. 
It measures track position in the 
transverse 
view with respect to the beam, 
which travels along the $z$ direction.
A five-layer silicon detector (SVX\ II) surrounding L00
measures track positions at
radii of 2.5 to 10.6 cm. 
Each of these layers
provides a transverse measurement
and a stereo measurement of $90^{\circ}$
(three layers) or  $\pm 1.2^{\circ}$ (two layers)
with respect to the beam direction.
The outermost silicon detector (ISL) lies between 19 cm and 30 cm radially, and
provides one or two track position measurements, 
depending on the track pseudorapidity ($\eta$), where  
$\eta \equiv -\ln ( \tan( \theta /2) )$, with $\theta$ 
being the angle between the particle
momentum and the proton beam direction.
An open-cell drift chamber (COT) 
completes the tracking system, and covers the radial region from
43 cm to 132 cm.
The COT consists of 96 sense-wire layers, arranged in 8 superlayers 
of 12 wires each.  Four of these superlayers provide axial measurements and four
provide  stereo views of $\pm2^{\circ}$.

Electromagnetic and hadronic calorimeters surround the solenoid coil.
 Muon candidates from the decay $J/\psi \rightarrow \mu^+ \mu^-$
 are  identified by  two sets of drift  chambers located radially
 outside  the     calorimeters.
The central muon chambers
cover the pseudorapidity region $ |\eta| <  0.6$, 
and  detect muons with 
transverse momentum $p_T > 1.4 $ GeV/$c$,
where the transverse momentum $p_T$ is defined as the component of the particle
momentum perpendicular to the proton  beam direction.
A second muon system covers the region $  0.6 < |\eta| <  1.0$ and
detects  muons with $p_T > 2.0 $ GeV/$c$.  
Muon selection
is based on matching these measurements to COT tracks, both in
projected position and angle.
The analysis presented here  is based on events recorded with a trigger
that is dedicated to the collection of a 
$J/\psi \rightarrow \mu^+ \mu^-$ sample.
The first level of the three-level trigger system requires two
muon candidates with matching tracks in the COT and muon chamber systems.
The second level imposes the requirement that muon candidates
have opposite charge 
and limits the accepted range of opening angle.  The highest level 
of the $J/\psi$  trigger reconstructs the muon pair in software, and 
requires that the invariant mass of the pair falls within the range
$2.7-4.0$ GeV/$c^2$.

The mass resolution and acceptance for the
$b-$hadrons used in this analysis are
studied with a Monte Carlo simulation that generates $b$ quarks
according to a next-to-leading-order calculation \cite{NDE},
and produces  events containing
final state hadrons by simulating
$b$ quark fragmentation \cite{Peterson}.
The final state decay processes are simulated with 
the {\sc EvtGen} \cite{EvtGen} decay program, 
a value of 6.12 GeV/$c^2$ is  taken
for the $\Omega^-_{b}$ mass, and all simulated $b$-hadrons are produced
without polarization.
The generated events are 
input to the detector and trigger simulation based on a 
{\sc GEANT3} description \cite{Geant}  
and processed through 
the same reconstruction and
analysis algorithms that are used for the data.

\section{Particle Reconstruction Methods \label{sect:Reconstruction}}

This analysis combines the trajectories of charged particles to
infer the presence of several different parent hadrons.
These hadrons are distinguished by their lifetimes, due to their
weak decay.  Consequently, it is useful to define two quantities
that are used frequently throughout the analysis which relate
the path of weakly decaying objects to their points of origin.
Both quantities are defined in the transverse view, and
make use of the point of closest approach, $\vec {r_{c}}$,
of the particle trajectory to a point of origin, and the measured
particle decay position,  $\vec {r_{d}}$.
The first quantity used here is transverse flight distance $f(h)$,
of hadron $h$, which is the distance
a particle has traveled in the transverse view.  
For neutral objects, flight distance is given by
$f(h) \equiv (\vec {r_{d}} - \vec {r_{c}})
 \cdot \vec{p_T}(h)/|\vec{p_T}(h)|$, where 
$\vec{p_T}(h)$ is the transverse momentum of the hadron candidate. 
For charged objects, the flight distance is calculated as the arc 
length in the transverse view from $\vec {r_{c}}$ to $\vec{r_{d}}$.
A complementary quantity used in this analysis is transverse impact distance $d(h)$,
which is the distance of the point of closest approach to the 
point of origin.  For neutral particles, transverse impact distance is given by
$d(h) \equiv  |(\vec{r_{d}} - \vec {r_{c}})
\times \vec{p_T}(h)|/|\vec{p_T}(h)|$.
The impact distance of charged particles is simply the distance from
$\vec {r_{c}}$ to the point of origin.
The measurement uncertainty on impact distance, $\sigma_{d(h)}$, is
calculated from the track parameter uncertainties and the
uncertainty on the point of origin. 

Several different selection criteria are employed in this analysis to
identify the particles used in $b-$hadron reconstruction.  These criteria 
are based on the resolution or acceptance of the CDF detector.  No optimization
procedure has been used to determine the exact value of any selection
requirement, since the analysis spans several final states and comparisons
between optimized selection requirements would necessarily be model-dependent.

\subsection{$J/\psi$ Reconstruction}
\begin{figure}[p]
\psfig{figure=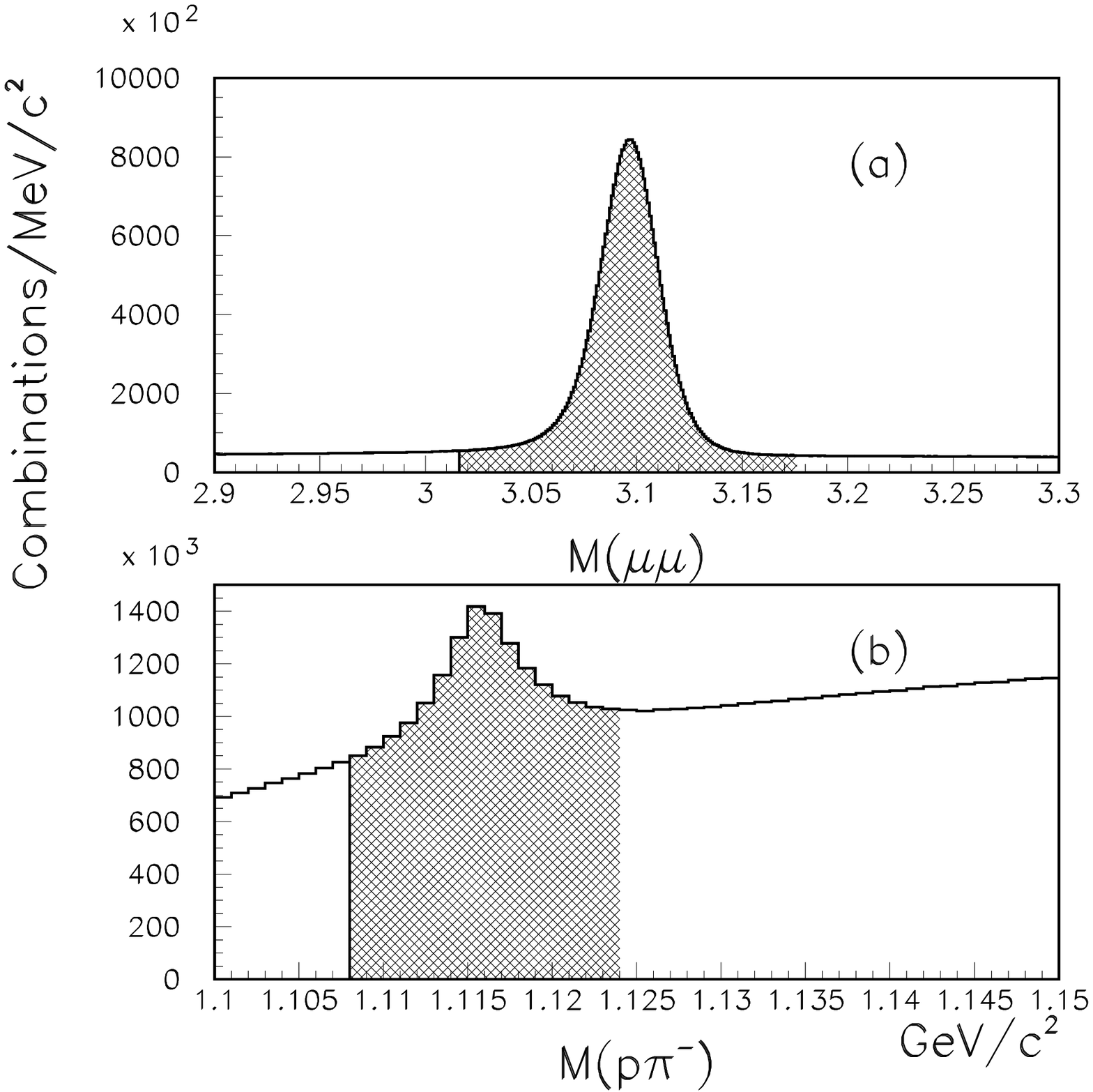, width=3.0in}
\caption{(a) The $\mu^+ \, \mu^-$ mass distribution obtained in 
an integrated luminosity of 4.2 fb$^{-1}$.  
The mass range used
for the $J/\psi$ sample is indicated by the shaded area.
(b) The $p \, \pi^-$ mass distribution obtained 
in events containing $J/\psi$ candidates.
The mass range used
for the $\Lambda$ sample is indicated by the shaded area.
\label{fig:prd_inclu_mumu} }
\end{figure}

The analysis of the data begins with a selection of well-measured
$J/\psi \rightarrow \mu^+ \mu^-$ candidates.
The trigger requirements are confirmed by selecting
 events that contain two
 oppositely  charged  muon candidates, each with matching 
 COT and muon chamber tracks.
Both muon tracks are required to have associated position
measurements in at least three layers of the SVX\ II and
a two-track invariant  mass within 80 MeV/$c^2$ of the
world-average  $J/\psi$ mass \cite{PDG}. 
This range was chosen for consistency with our earlier $b-$hadron mass
measurements \cite{CDF_Bmass}.
The $\mu^+ \, \mu^-$ mass distribution obtained in these data
is shown in Fig.\ \ref{fig:prd_inclu_mumu}(a).
This data sample provides approximately $2.9\times 10^7$  $J/\psi$
candidates, measured with an average mass resolution of $\sim20$ MeV/$c^2$.
 
\subsection{Neutral Hadron Reconstruction}

The reconstruction of $K^0_s$, $K^{*}(892)^{0}$, and $\Lambda$
candidates uses all tracks with $p_T \, >$ 0.4 GeV/$c$
found in the COT, that are not associated with muons in the $J/\psi$ 
reconstruction.  
Pairs of oppositely charged tracks are combined to identify these
neutral decay candidates, and silicon detector information
is not used. 
Candidate selection for these neutral states is 
based upon the mass calculated for each track pair,
which is required to fall within the ranges given 
in Table\ \ref{table:mass_range}
after the appropriate mass assignment for each track.

Candidates for $K^0_s$ decay are chosen by assigning the pion
mass to each track, and mass is measured with a resolution of
$\sim6$ MeV/$c^2$.
Track pairs used for $K^{*}(892)^{0} \rightarrow K^+ \pi^-$ candidates have both
mass assignments examined.
A broad mass selection range is chosen for the $K^{*}(892)^{0}$ signal, 
due to its 
natural width of 50 MeV/$c^2$ \cite{PDG}.
In the situation where both assignments fall 
within our selection mass range, only the combination closest to the nominal 
$K^{*}(892)^{0}$ mass is used.
For $\Lambda \rightarrow p \, \pi^-$ candidates, 
the proton (pion) mass is assigned 
to the track with the higher (lower) momentum.
This mass assignment is 
always correct for the $\Lambda$ candidates 
used in this analysis
because of the
kinematics of $\Lambda$ decay and the lower limit
in the transverse momentum
acceptance of the tracking system. 
Backgrounds to the $K^0_s$ ($c\tau=2.7$ cm) and  
$\Lambda$ ($c\tau=7.9$ cm) \cite{PDG} are reduced by requiring
the flight distance 
of the $K^0_s$ and $\Lambda$ with respect to the primary vertex
(defined as the beam position in the transverse view) to be
greater than 1.0 cm, which corresponds to $\sim0.6 \, \sigma$.
The mass distribution of the $p \, \pi^-$ combinations 
with $p_T(\Lambda) \, > \, 2.0$ GeV/$c$
is plotted in Fig.\ \ref{fig:prd_inclu_mumu}(b), and contains approximately
$3.6\times 10^6$ $\Lambda$ candidates.

\subsection{Charged Hyperon Reconstruction}

For events that contain a $\Lambda$ candidate,
the remaining tracks reconstructed in the COT, again without additional
silicon information,
are assigned the pion or kaon mass, and
$\Lambda \, \pi^-$ or $\Lambda \, K^-$ combinations 
are identified that are consistent with the decay process
$\Xi^- \rightarrow \Lambda \, \pi^-$ or 
$\Omega^- \rightarrow \Lambda \, K^-$.  
Analysis of the simulated $\Xi^-_{b}$ events
shows that the $p_T$ distribution of the $\pi^-$ daughters of
reconstructed $\Lambda$ and $\Xi^-$ decays 
falls steeply with increasing $p_T(\pi^-)$.
Consequently, tracks with $p_T$ 
as low as 0.4 GeV/$c$ are used for these reconstructions.
The simulation
also indicates that the $p_T$ distribution of the $K^-$ daughters
from $\Omega^-$ decay has a higher average value, and declines with $p_T$
much more slowly, than the $p_T$ distribution of the
pions from $\Lambda$ or $\Xi^-$ decays.
A study of the $\Lambda \, K^-$ 
combinatorial backgrounds in two 8 MeV/$c^2$ mass
ranges and centered $\pm20$ MeV/$c^2$ from the
$\Omega^-$ mass indicates that the background track $p_T$
distribution is also steeper than the expected 
distribution of $K^-$ from $\Omega^-$ decay.
Therefore, $p_T(K^-) > 1.0$ GeV/$c$ is required
for our $\Omega^-$ sample, which reduces the combinatorial background
by 60\%, while reducing the $\Omega^-$ signal predicted by our Monte Carlo 
simulation by 25\%.

\begin{figure}[p]
\psfig{figure=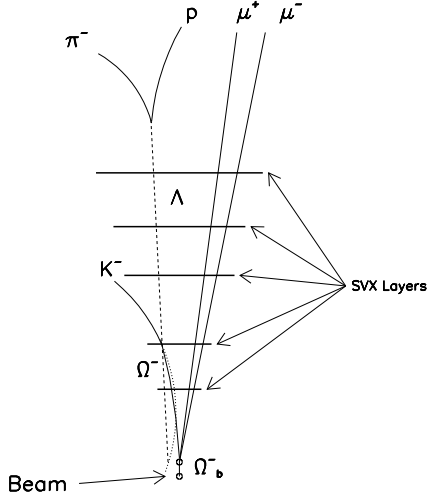, width=3.0in}
\caption{An illustration (not to scale) of the 
$\Omega^-_b \rightarrow J/\psi \, \Omega^-$,
$J/\psi \rightarrow \mu^+\, \mu^-$, 
$\Omega^- \rightarrow \Lambda \, K^-$,
and $\Lambda \rightarrow p \, \pi^-$ final state as seen in the
view transverse to the beam direction.  
Five charged tracks are used to identify three decay vertices.
The final fit of these track trajectories constrains
the decay hadrons ($J/\psi, \, \Omega^-,$ and $\Lambda$) to their
nominal masses and the helix of the $\Omega^-$ to originate from
the $J/\psi$ decay vertex.  
The trajectory of the $K^-$ 
is projected back, indicated by a dotted curve, to illustrate 
how an alternative, incorrect intersection with the $\Lambda$ 
trajectory could exist.  A comparison of the fit quality of the two 
$\Lambda \, K^-$ intersections is used to choose a preferred solution.
\label{fig:cartoon} }
\end{figure}

An illustration of the full $\Omega^-_b$ final state that is
reconstructed in this analysis is shown in Fig.\ \ref{fig:cartoon}.
Several features of the track topology are used to reduce the
background to this process.
In order to obtain the best possible 
 mass resolution for $\Xi^-$ and $\Omega^-$ candidates, 
the reconstruction requires a convergent  
fit of the three tracks that simultaneously 
constrains the $\Lambda$
decay products to the $\Lambda$ mass, and the
$\Lambda$ trajectory to intersect with the helix of 
the $\pi^-(K^-)$ originating from the $\Xi^-(\Omega^-)$ candidate.
In addition,
the flight distance of the $\Lambda$ candidate with respect to the
reconstructed decay vertex of the $\Xi^-(\Omega^-)$ candidate
is required to exceed 1.0 cm.
Similarly, due to the long lifetimes of the $\Xi^-$ ($c\tau$ = 4.9 cm)
and $\Omega^-$ ($c\tau$ = 2.5 cm) \cite{PDG}, a flight distance
of at least 1.0 cm (corresponding to $\sim 1.0 \, \sigma$) 
with respect to the primary vertex
is required.  This requirement removes $\sim75\%$ of the background 
to these long-lived
particles, due to prompt particle production.

\begin{figure}[p]
\psfig{figure=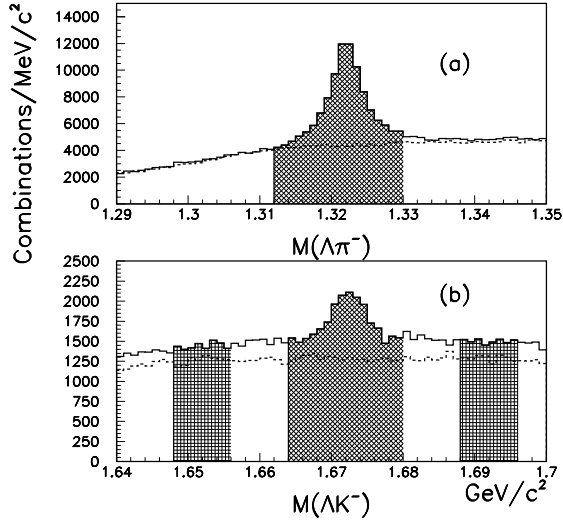, width=3.0in}
\caption{The invariant mass distributions of 
(a) $\Lambda \, \pi^-$ combinations and
(b) $\Lambda \, K^-$ combinations 
in events containing $J/\psi$ candidates.
Shaded areas indicate the mass ranges used for $\Xi^-$ and $\Omega^-$ 
candidates. 
The dashed histograms in each distribution correspond to 
$\Lambda \, \pi^+$(a) and $\Lambda \, K^+$(b) combinations.
Additional shading in (b) correspond to sideband regions discussed in 
Section\ \ref{sect:signif}.
\label{fig:prd_inclu_xi} }
\end{figure}

Possible kinematic reflections are removed from the $\Omega^-$ sample
by requiring that the combinations in the sample fall outside the 
$\Xi^-$ mass range listed in Table\ \ref{table:mass_range} when the candidate
$K^-$ track is assigned the mass of the $\pi^-$.  
In some instances, the rotation of the $\pi^-(K^-)$ helix produces a situation
where two
$\Lambda \, \pi^-(\Lambda \, K^-)$ 
vertices satisfy the constrained fit and displacement
requirements. These situations are resolved with the tracking
measurements in the longitudinal view.
The candidate with the poorer value of probability $P(\chi^2)$ for
the $\Xi^-(\Omega^-)$ fit is dropped from the sample.
An example of such a combination is illustrated in Fig.\ \ref{fig:cartoon}.
The complexity of the $\Xi^-(\Omega^-)$ and $\Lambda$ decays allows for
occasional combinations where the proper identity of the three tracks
is ambiguous.  An example is where the $\pi^-(K^-)$ candidate track and the
$\pi^-$ 
candidate track from $\Lambda$ decay are interchanged, and the interchanged
solution satisfies the various mass and flight distance requirements.
A single, preferred candidate is chosen by retaining only
the fit combination with the highest $P(\chi^2)$ of all the possibilities.
Requiring the impact distance with respect to the primary vertex to be less than
$3\sigma_{d(h)}$ and $p_T(h) \, > \, 2.0$ GeV/$c$
results in the combinations 
shown in the $\Lambda \pi^-$
and $\Lambda K^-$ mass distributions of Fig.\ \ref{fig:prd_inclu_xi}.
Approximately $41 \, 000$ $\Xi^-$ and $3500$ $\Omega^-$ candidates
are found in this data sample.

\begin{figure}[p]
\psfig{figure=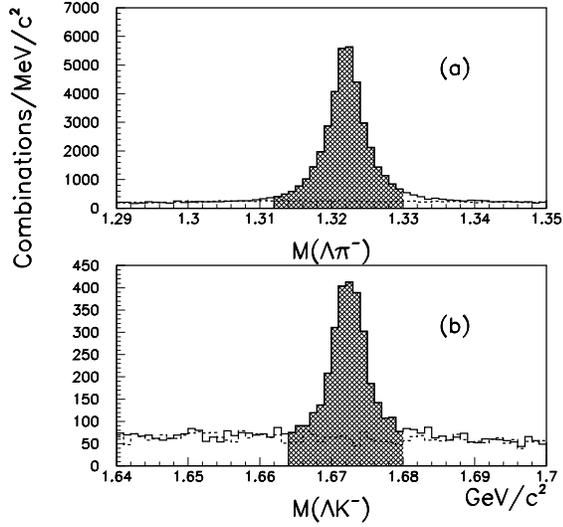, width=3.0in}
\caption{The invariant mass distributions of 
(a) $\Lambda \, \pi^-$ combinations and
(b) $\Lambda \, K^-$ combinations 
in events containing $J/\psi$ candidates.
These combinations require 
silicon information to be used on the hyperon track
and the impact distance with respect to the primary vertex must not exceed three
times its measurement resolution.
Shaded areas indicate the mass ranges used for $\Xi^-$ and $\Omega^-$ 
candidates. 
The dashed histograms in each distribution correspond to 
$\Lambda \, \pi^+$ and $\Lambda \, K^+$ combinations.
\label{fig:prd_inclu_xi_sil} }
\end{figure}

The mass distributions in Fig.\ \ref{fig:prd_inclu_xi} 
show  clear $\Xi^-$ and $\Omega^-$ signals.  However, the $\Omega^-$ 
signal has a substantially larger combinatorial background.
The kinematics of hyperon decay and the lower $p_T$ limit of 0.4 GeV/$c$
on the decay daughter tracks force the majority of charged hyperon 
candidates to have $p_T >$ 1.5 GeV/$c$.  
This fact, along with the long lifetimes of the $\Xi^-$ 
and $\Omega^-$,
results in a significant fraction of the hyperon candidates having
decay vertices located several centimeters radially outward from the 
beam position.
Therefore, we are able to
refine the charged hyperon  reconstruction by making use of the 
improved determination of the trajectory that can be obtained by tracking 
these particles in
the silicon detector.  The charged hyperon candidates have an additional
fit performed with the three tracks  that simultaneously 
constrains
both the $\Lambda$ and $\Xi^-$ or $\Omega^-$  masses of the 
appropriate track combinations,
and provides the best possible estimate of the hyperon momentum and decay
position.
The result of this fit is used to define a helix that
serves as the seed for an algorithm that associates silicon detector
hits with the charged hyperon track.  
Charged hyperon
candidates with track measurements in at least one layer of the 
silicon detector
have excellent impact distance resolution (average of 60 $\mu$m) for the
charged hyperon track. 
The mass distributions for the subset of the inclusive $\Lambda \pi^-$
and $\Lambda K^-$ combinations which are found in the silicon detector, and 
have an impact distance with respect to the primary vertex
$d(h) < 3\sigma_{d(h)}$ are shown in Fig.\ \ref{fig:prd_inclu_xi_sil}.
This selection provides approximately $34 \, 700$ $\Xi^-$ candidates
and $1900$ $\Omega^-$ candidates 
with very low combinatorial background, which allows us to confirm
the mass resolutions used to select hyperons.  Unfortunately,
the shorter lifetime of the  $\Omega^-$ makes the silicon 
selection less efficient
than it is for the $\Xi^-$.  Therefore, silicon detector information on the
hyperon track is used whenever
it is available, but is not imposed as a requirement for the
$\Omega^-$ selection.

\subsection{$b$-Hadron Reconstruction}

The reconstruction of $b$-hadron candidates uses the same method for
each of the states reconstructed for this analysis.
The $K$ and hyperon candidates are 
combined with the $J/\psi$ candidates by fitting
the full four-track or five-track state with constraints appropriate
for each decay topology and intermediate hadron state.  
Specifically, the $\mu^+ \, \mu^-$ mass is constrained
to the nominal $J/\psi$ mass \cite{PDG}, 
and the neutral $K$ or hyperon candidate
is constrained to originate from the $J/\psi$ decay vertex.
In addition, the fits that include the
charged hyperons constrain the $\Lambda$ candidate tracks to the nominal
$\Lambda$ mass \cite{PDG}, and the $\Xi^-$ and $\Omega^-$ candidates to their
respective nominal masses \cite{PDG}.
The $\Xi^-_{b}$ and $\Omega_b^-$
 mass resolutions obtained from simulated events
are found to be approximately $12$ MeV/$c^2$, a value that is
comparable to
the mass resolution obtained with the CDF II detector for other $b$-hadrons
 with a $J/\psi$ meson in the final state \cite{CDF_Bmass}.

The selection used to reconstruct $b$-hadrons is 
chosen to be as generally applicable as possible, in order to minimize
systematic effects in rate comparisons, and to provide confidence that the
observation of $\Omega^-_{b} \rightarrow$ $J/\psi \, \Omega^-$
is not an artifact of the selection.
Therefore, the final samples
of all $b$-hadrons used in this analysis
are selected with a small number of requirements
that can be applied to any $b$-hadron candidate.  First, 
$b$-hadron candidates are required 
to have $p_T \, > $ 6.0 GeV/$c$ and the neutral 
$K$ or hyperon to have $p_T \, > $ 2.0 GeV/$c$.  
These $p_T$ requirements restrict the sample to candidates
that are within the kinematic range where our acceptance is well modeled.
Mass ranges are 
imposed on the decay products of the $K$ and hyperon candidates 
based on observed mass resolution or natural width, 
as listed in Table\ \ref{table:mass_range}.
The promptly-produced combinatorial background is suppressed by rejecting 
candidates with low proper decay time, 
$t \equiv f(B) M(B)/(c \, p_T(B))$, 
where $M(B)$ is the measured mass,
$p_T(B)$ is the transverse momentum, and
$f(B)$ is the flight distance of the $b$-hadron candidate
measured with respect to the primary vertex. 

Combinations that are inconsistent with having
originated from the collision are rejected by imposing an upper limit on the
impact distance of the $b$-hadron candidate measured with respect to the
primary vertex $d_{PV}$.  
Similarly, the trajectory of the decay hadron is required to originate from the
$b$-hadron decay vertex by imposing an upper limit on its impact distance
$d_{\mu\mu}$ 
with respect to the vertex found in the 
$J/\psi$ fit.
These two impact distance quantities are compared to their 
measurement uncertainties
$\sigma_{d_{PV}}$ and $\sigma_{d_{\mu\mu}}$ when they are used.

\begin{table}[ht]
\begin{center}
\caption{Mass ranges around the nominal mass value \cite{PDG}
used for the $b$-hadron decay products.
\label{table:mass_range}}
\begin{tabular}{cc}
\hline \hline
Resonance (Final State) & Mass range (MeV/$c^2$)\\
\hline
 $J/\psi\, (\mu^+ \mu^-)$    &  $\pm80$ \\
 $K^{*}(892)^{0} \, (K^+ \pi^-)$     &  $\pm30$ \\
 $K^0_s \, (\pi^+ \pi^-)$    &  $\pm20$ \\
 $\Lambda \, (p \pi^-)$      &  $\pm9$ \\
 $\Xi^- \, (\Lambda \pi^-)$  &  $\pm9$ \\
 $\Omega^- \, (\Lambda K^-)$ &  $\pm8$ \\
\hline \hline
\end{tabular}
\end{center}
\end{table}

\section{Observation of the Decay 
$\Omega^-_b\rightarrow$ $J/\psi \, \Omega^-$
 \label{sect:signif}}

\begin{figure}[p]
\psfig{figure=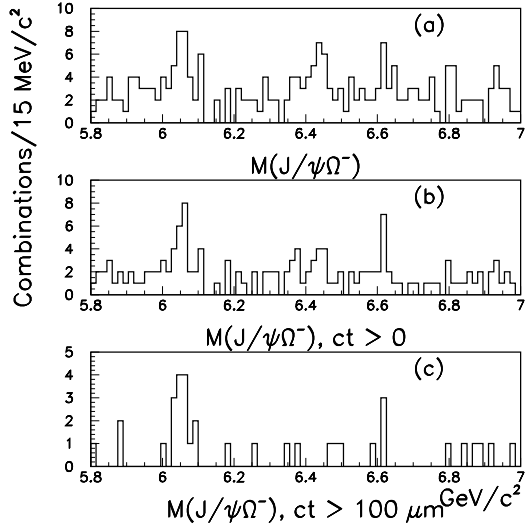, width=3.0in}
\caption{(a) The mass distribution of all
$J/\psi \, \Omega^-$ combinations.
(b) The $J/\psi \, \Omega^-$  mass distribution for candidates with
$ct \, > \, 0$.  This requirement removes half of the combinations 
due to prompt production.
(c) The $J/\psi \, \Omega^-$  mass distribution for candidates with
$ct \, > \, 100 \, \mu$m.  This requirement removes nearly all 
combinations directly produced in the $p\bar p$ collision.
 \label{fig:prd_omegab_sel} }
\end{figure}

\begin{figure}[p]
\psfig{figure=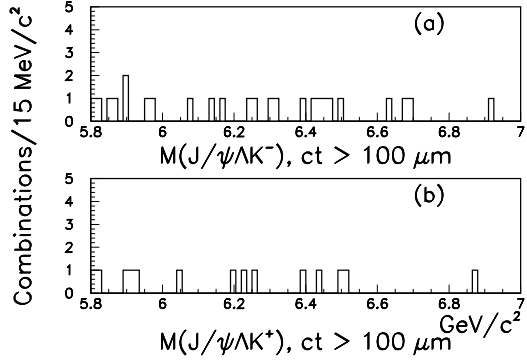, width=3.0in}
\caption{(a)  The invariant mass distributions of 
$J/\psi \, \Lambda K^-$ combinations for candidates with
$\Lambda K^-$ in the $\Omega^-$ sidebands.
(b) The invariant mass distributions of 
$J/\psi \, \Lambda K^+$ combinations for candidates with
$\Lambda K^+$ in the $\Omega^-$ signal range.
All other selection requirements are as in Fig.\ \ref{fig:prd_omegab_sel}(c).
\label{fig:prd_omegab_side} }
\end{figure}

The $J/\psi \, \Omega^-$ mass distribution
with $d_{PV} \, < \,3\sigma_{d_{PV}}$ and
$d_{\mu\mu} \, < \,3\sigma_{d_{\mu\mu}}$
is shown in Fig.\ \ref{fig:prd_omegab_sel} 
for the full sample and two different requirements of $ct$. 
The samples with a $ct$ requirement of 100 $\mu$m or greater
show clear evidence of a resonance near a mass of 6.05 GeV/$c^2$,
with a width consistent with our measurement resolution.
Mass sideband regions have been defined as 8 MeV/$c^2$ wide ranges,
centered 20 MeV/$c^2$ above and below the nominal $\Omega^-$ mass, as 
indicated in Fig.\ \ref{fig:prd_inclu_xi}.
The $J/\psi \, \Lambda K^-$ mass distribution for combinations that
populate the $\Omega^-$ mass sideband regions is shown in 
Fig.\ \ref{fig:prd_omegab_side}(a).
In addition, the $J/\psi \, \Lambda K^+$ distribution for combinations
where the $\Lambda K^+$ mass populates the $\Omega^-$ signal region
is shown in Fig.\ \ref{fig:prd_omegab_side}(b).  No evidence of any
mass resonance structure appears in either of these distributions.

\begin{figure}[p]
\psfig{figure=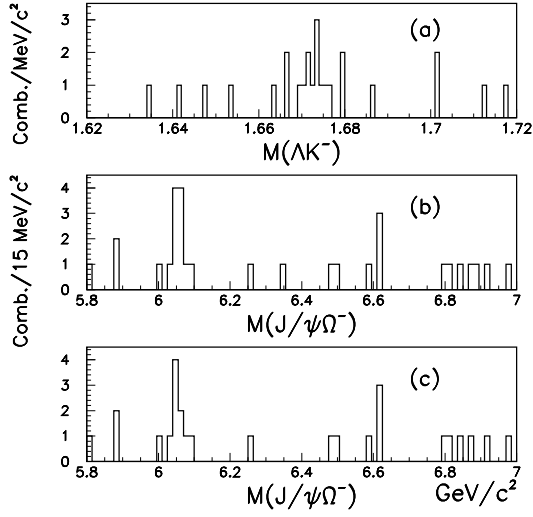, width=3.0in}
\caption{(a)  The invariant mass distribution of 
$\Lambda \, K^-$ combinations for candidates with
$J/\psi \Lambda K^-$ masses in the range 6.0 - 6.1 GeV/$c^2$.
(b) The invariant mass distribution of 
$J/\psi \, \Lambda K^-$ combinations for candidates with
$\Lambda K^-$ masses within 6 MeV/$c^2$ of the $\Omega^-$ mass.
(c) The invariant mass distributions of 
$J/\psi \, \Lambda K^-$ combinations for candidates with
$\Lambda K^-$ masses within 4 MeV/$c^2$ of the $\Omega^-$ mass.
All other selection requirements are as in Fig.\ \ref{fig:prd_omegab_sel}(c).
\label{fig:prd_omegab_cuts_1} }
\end{figure}

The only selection criteria unique to this analysis are those used
in the $\Omega^-$ selection.  
Therefore, the quantities used in the  $\Omega^-$ selection were 
varied to provide confidence that the resonance structure centered at
6.05 GeV/$c^2$ is not peculiar to the values
of the selection requirements that were chosen.
The first selection criterion that was varied is the
$\Lambda \, K^-$ mass range used to define the $\Omega^-$ sample.
For the candidates that satisfy the selection used in
Fig.\ \ref{fig:prd_omegab_sel}(c), the $\Lambda \, K^-$ mass range
was opened to $\pm50$ MeV/$c^2$.  The $\Lambda \, K^-$ 
mass distribution for combinations with
a $J/\psi \, \Lambda K^-$ mass in the range $6.0-6.1$ GeV/$c^2$
is shown in Fig \ \ref{fig:prd_omegab_cuts_1}(a).
A clear indication of an $\Omega^-$ signal can be seen, as expected for a real
decay process.
The $\Lambda \, K^-$ mass range of $\pm8$ MeV/$c^2$  used in the selection
was chosen to be inclusive for all likely $\Omega^-$ candidates.
More restrictive mass ranges for the $\Omega^-$ selection are shown in
Fig. \ \ref{fig:prd_omegab_cuts_1}(b) and  
Fig. \ \ref{fig:prd_omegab_cuts_1}(c), where the $\Lambda \, K^-$ mass range
is reduced to $\pm6$ and $\pm4$ MeV/$c^2$, respectively.
The apparent excess of $J/\psi \, \Omega^-$ 
combinations in the $6.0-6.1$ GeV/$c^2$
mass range is retained for these more restrictive requirements.

\begin{figure}[p]
\psfig{figure=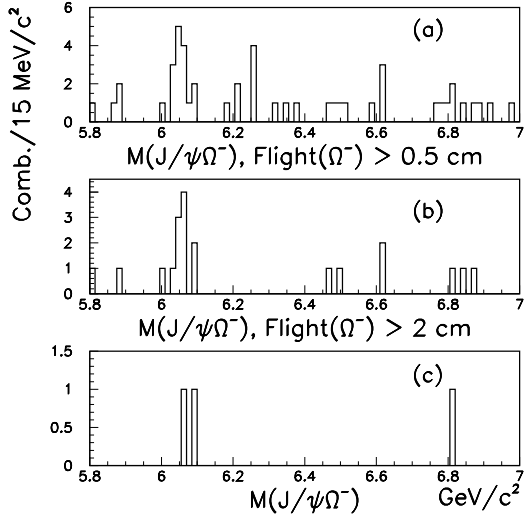, width=3.0in}
\caption{
(a,b) The invariant mass distribution of 
$J/\psi \, \Omega^-$ combinations for candidates where the
transverse flight requirement of the $\Omega^-$ is greater than 0.5 cm
and 2.0 cm. 
(c) The invariant mass distribution of 
$J/\psi \, \Omega^-$ combinations for candidates with at least one SVXII 
measurement on the $\Omega^-$ track.
All other selection requirements are as in Fig.\ \ref{fig:prd_omegab_sel}(c).
\label{fig:prd_omegab_cuts_2} }
\end{figure}

A transverse flight requirement of 1 cm is used for the $\Omega^-$ selection.
A lower value allows more promptly produced background into the sample,
due to our measurement resolution.  A higher value reduces 
our acceptance, due to the decay of the $\Omega^-$.  Two variations
of the flight requirement are shown in 
Fig.\ \ref{fig:prd_omegab_cuts_2}(a) and 
Fig. \ \ref{fig:prd_omegab_cuts_2}(b).  No striking changes
in the $J/\psi \, \Omega^-$ mass distribution 
appear for these variations.  A more restrictive flight cut
can also be imposed, which limits the sample to $\Omega^-$ candidates
that are measured in the SVXII (inner radius is 2.5 cm), 
and provides the extremely pure
$\Omega^-$ sample seen in Fig. \ \ref{fig:prd_inclu_xi_sil}. 
Two candidates in the $6.0-6.1$ GeV/$c^2$
mass range are retained, and no others in the range expected for the
$\Omega_b^-$.

\begin{figure}[p]
\psfig{figure=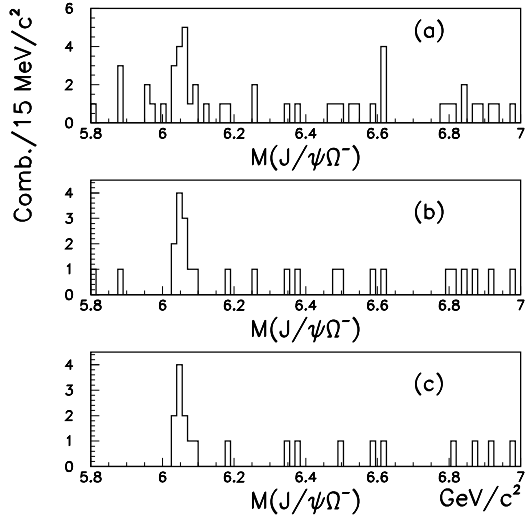, width=3.0in}
\caption{The invariant mass distributions of 
$J/\psi \, \Omega^-$ combinations for candidates with
three alternative requirements for the transverse momentum of the $K^-$.
(a) $p_T(K^-) > 0.8$ GeV/$c$.
(b) $p_T(K^-) > 1.2$ GeV/$c$.
(c) $p_T(K^-) > 1.4$ GeV/$c$.
All other selection requirements are as in Fig.\ \ref{fig:prd_omegab_sel}(c).
\label{fig:prd_omegab_cuts_3} }
\end{figure}

A $p_T(K^-) > 1.0$ GeV/$c$ requirement is used in the $\Omega^-$ selection,
to reduce the background due to tracks from fragmentation and other
sources.  The effect of three different selection values is shown in
Fig \ \ref{fig:prd_omegab_cuts_3}.  The excess of
$J/\psi \, \Omega^-$ combinations in the
mass range $6.0-6.1$ GeV/$c^2$ appears for all
$p_T(K^-)$ values shown, and is probably a higher fraction of
the total combinations seen for the more restrictive requirements.
We conclude that the excess of $J/\psi \, \Omega^-$ combinations 
near 6.05 GeV/$c^2$ is not an artifact 
of our selection process.  

The mass, yield, and significance of the resonance candidate
in Fig.\ \ref{fig:prd_omegab_sel}(c) are obtained by performing 
an unbinned likelihood fit on the mass distribution of candidates.
The likelihood function that is maximized has the form
\begin{eqnarray}
{\cal L} & = & \prod_i^N (f_s{\cal P}^s_i + (1-f_s) {\cal P}^b_i) \nonumber \\
 & = & \prod_i^N \left ( 
f_sG(m_i,m_0,s_m\sigma^m_i) + (1-f_s) P^n(m_i)\right ),
\label{eq:unbinned}
\end{eqnarray}
where $N$ is the number of candidates in the sample,
${\cal P}^s_i$ and ${\cal P}^b_i$ are 
the probability distribution functions for the signal and background, 
respectively,
$G(m_i,m_0,s_m\sigma^m_i)$ is a Gaussian distribution with average
$m_0$ and characteristic width $s_m\sigma^m_i$ to describe the signal,
$m_i$ is the mass obtained for a single $J/\psi \Omega^-$ candidate, 
$\sigma^m_i$ is the resolution on that mass,
and $P^n(m_i)$ is a polynomial of order $n$.  
The quantities
obtained from the fitting procedure include 
$f_s$, the fraction of the candidates identified as signal, 
$m_0$, the best average mass value, 
$s_m$, a scale factor on the mass resolution, 
and the coefficients of $P^n(m_i)$.

Two applications of this mass  fit are
used with the $J/\psi \, \Omega^-$ combinations shown in 
Fig.\ \ref{fig:prd_omegab_sel}c.  For this data sample,
all background polynomials are first order and
the mass resolution is fixed to 12 MeV/$c^2$.
The first of these fits allows the remaining parameters to vary.
The second application corresponds to the null signal hypothesis, and fixes
$f_s=0.0$, thereby removing $f_s$ and $m_0$ as fitting variables.
The value of $-2\ln{\cal L}$ for the null hypothesis
exceeds the fit with variable $f_s$ by 27.9 units for the sample with  
$ct \, > \, 100 \, \mu$m.  We interpret this as 
equivalent to a $\chi^2$ with two degrees of freedom
(one each for $f_s$ and $m_0$), whose 
probability of occurrence is
$8.7\times10^{-7}$, corresponding to a $4.9\sigma$ 
significance.  
This calculation was checked by a second technique, which used a simulation
to estimate the probability for a pure background sample to
produce the observed signal anywhere within a 400 MeV/$c^2$ range.  
The simulation randomly distributed the number of entries in 
Fig.\ \ref{fig:prd_omegab_sel}c over its mass range.  
Each resulting distribution was then fit with both the null hypothesis and
where $f_s$ and $m_0$ are allowed to vary.
The simulation result, based on the distribution of $\delta (2\ln{\cal L})$
from $10^7$ trials,
confirmed the significance obtained by the ratio-of-likelihoods test.

An alternative to the mass fit obtained by maximizing Eq. (\ref{eq:unbinned})
is to simultaneously fit mass and lifetime information.
This can be accomplished by replacing the probability distribution
functions used in the likelihood definition.  
Lifetime information for the signal term can be added by setting
${\cal P}^s_i={\cal P}^{s,m}_i {\cal P}^{s,ct}_i$
where ${\cal P}^{s,m}_i$ is the mass distribution as in Eq. (\ref{eq:unbinned}),
and ${\cal P}^{s,ct}_i$ describes the distribution in $ct$.
The background can have both prompt and $b$-hadron decay contributions.
These are included by setting
${\cal P}^b_i=(1-f_B){\cal P}^{p,m}_i{\cal P}^{p,ct}_i +
f_B{\cal P}^{B,m}_i{\cal P}^{B,ct}_i$ 
where ${\cal P}^{p,m}_i$ and ${\cal P}^{p,ct}_i$ are the prompt mass 
and lifetime terms, 
${\cal P}^{B,m}_i$ and ${\cal P}^{B,ct}_i$ are the $b$-hadron decay
terms, and $f_B$ is the fraction of the background due to $b$-hadron decay.
The time distribution of the prompt background ${\cal P}^{p,ct}_i$
is simply due to measurement
resolution and is given by $G(ct_i,0,\sigma^{ct}_i)$, 
where $ct_i$ is the $ct$ of candidate $i$, and $\sigma^{ct}_i$
is its measurement resolution.
The time probability  distribution of the 
signal is an exponential convoluted with the measurement resolution,
given by
\begin{eqnarray}
\lefteqn{{\cal S}(ct_i,c\tau,\sigma^{ct}_i) =} \nonumber \\
 & & \frac{1}{c\tau}\exp \left( \frac{1}{2} \left( \frac{\sigma^{ct}_i}{c\tau}
\right)^2
- \frac{ct_i}{c\tau} \right) 
\textrm{erfc} \left( \frac{\sigma^{ct}_i}{\sqrt{2}c\tau}-
\frac{ct_i}{\sqrt{2}\sigma^{ct}_i}\right),
\label{eq:smear}
\end{eqnarray}
where $\tau$ is the $b$-hadron lifetime.
A similar model is used for the $b$-hadron decay background.
Therefore, these time distributions are given by
${\cal P}^{s,ct}_i={\cal S}(ct_i,c\tau,\sigma^{ct}_i)$ and
${\cal P}^{B,ct}_i={\cal S}(ct_i,c\tau_B,\sigma^{ct}_i)$, 
and the new likelihood becomes
\begin{eqnarray}
{\cal L} & = &\prod_i^N \left ( f_s{\cal P}^{s,m}_i{\cal P}^{s,ct}_i + 
\right. \nonumber \\
& & (1-f_s) ((1-f_B){\cal P}^{p,m}_i {\cal P}^{p,ct}_i + \nonumber \\
& &\left. f_B {\cal P}^{B,m}_i {\cal P}^{B,ct}_i ) \right ).
\label{eq:unbinned_life}
\end{eqnarray}
 
The simultaneous mass and lifetime likelihood in Eq. (\ref{eq:unbinned_life})
is maximized for two different conditions.  Both calculations use 
$\sigma^m_i$ = 12 MeV/$c$, and $\sigma^{ct}_i$ = 30 $\mu$m, which 
is the average resolution found for all other final states 
reconstructed in this analysis. 
The first maximization allows all other parameters to
vary in the fit.  The second calculation fixes $f_s=0.0$, as 
was done for the mass fit.  
The value of $-2\ln{\cal L}$ obtained for the null hypothesis is
higher than the value obtained for the fully varying calculation by 37.3
units.  
We interpret this as equivalent to a $\chi^2$ with
three degrees of freedom, which has a probability of occurrence of
$4.0\times10^{-8}$, or a $5.5\sigma$ fluctuation.
Consequently, we 
interpret the  $J/\psi \, \Omega^-$ 
mass distributions shown in Fig.\ \ref{fig:prd_omegab_sel} 
to be the 
observation of a weakly decaying resonance, with a width consistent with the 
detector resolution.
We treat this resonance as observation of the  $\Omega^-_b$ baryon
through the decay process $\Omega^-_b \rightarrow J/\psi \, \Omega^-$.

\section{$\Xi^-_b$ and $\Omega^-_b$ Property Measurements
\label{sect:Properties}}
For the measurement of $\Omega^-_b$ properties, the impact distance 
requirements placed on the $J/\psi \, \Omega^-$ sample discussed above are
not used.  These requirements reduce the combinatorial background
to the $\Omega^-_b$ signal, but do not have the same efficiency for other
$b$-hadrons, since the silicon detector efficiency for the charged hyperons 
is different for each state.  
Consequently, the charged hyperon helix with silicon detector measurements
is not used any further.
The remainder of the analysis uses silicon information only 
on the muons of the final states.  The hadron tracks are all 
measured exclusively in the COT to achieve uniformity across all the
$b$-hadron states discussed in this paper.

\subsection{Mass Measurements}
\begin{figure}[p]
\psfig{figure=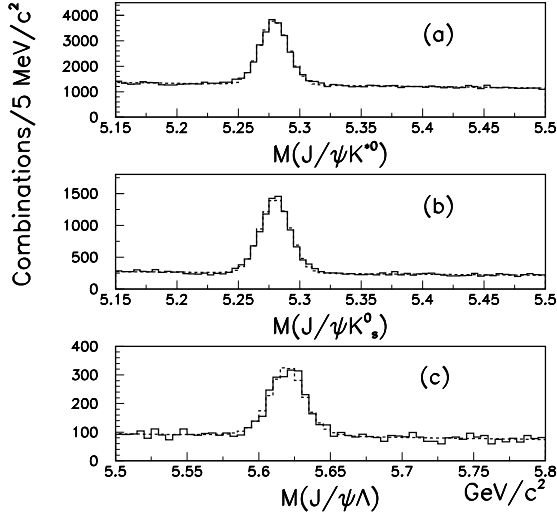, width=3.0in}
\caption{The invariant mass distributions of 
(a) $J/\psi \, K^{*}(892)^{0}$, (b) $J/\psi \, K^0_s$, and 
(c) $J/\psi \, \Lambda$ combinations for candidates with
$ct > 100 \, \mu$m.  
The projections of the unbinned mass fits are indicated by the 
dashed histograms.
\label{fig:prd_b0_mass} }
\end{figure}

\begin{figure}[p]
\psfig{figure=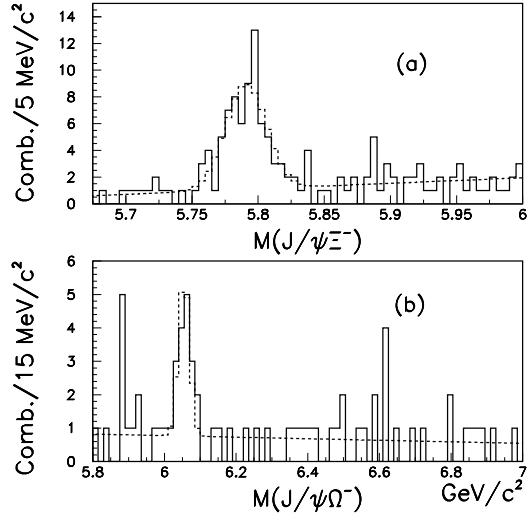, width=3.0in}
\caption{The invariant mass distributions of 
(a) $J/\psi \, \Xi^-$  and
(b) $J/\psi \, \Omega^-$ combinations for candidates with
$ct > 100 \, \mu$m.
The projections of the unbinned mass fit are indicated by the 
dashed histograms.
\label{fig:prd_xib_mass} }
\end{figure}

To reduce the background to $b$-hadrons due to prompt production, 
a $ct \, > \, 100 \, \mu$m requirement is placed on all
candidates for inclusion in the mass measurements.
Masses are calculated by maximizing the
likelihood function given in Eq. (\ref{eq:unbinned}).
The mass distributions of the candidates are shown in 
Figs.\ \ref{fig:prd_b0_mass} and \ref{fig:prd_xib_mass}, along with 
projections of the fit function.
The results of this fit 
are listed in Table\ \ref{table:mass}.
The resolution scale factor used for the $\Omega^-_b$
fit is fixed  to the value obtained from the $\Xi^-_b$,
since the small sample size makes a scale factor 
calculation unreliable.

\begin{table*}[ht]
\begin{center}
\caption{Masses obtained for $b$-hadrons.
\label{table:mass}}
\begin{tabular}{cccc}
\hline \hline
Resonance  & Candidates &Mass  (MeV/$c^2$) & Resolution Scale\\
\hline
 $B^0 (J/\psi \, K^{*}(892)^{0})$ & $15181\pm200$ &  $5279.2\pm0.2$ & $0.98\pm0.02$ \\
 $B^0 (J/\psi \, K^0_s) $ &  $7424\pm113$ &  $5280.2\pm0.2$ & $1.04\pm0.02$ \\
 $\Lambda^0_b$              &  $1509\pm58$  &  $5620.3\pm0.5$ & $1.04\pm0.02$ \\
 $\Xi^-_b $               &    $61\pm10$  &  $5790.9\pm2.6$ & $1.3\pm0.2$  \\
 $\Omega^-_b$             &    $12\pm4$   &  $6054.4\pm6.8$ & $1.3$ \\
\hline \hline
\end{tabular}
\end{center}
\end{table*}

The mass difference between the $B^0$ as measured in the $J/\psi \, K_s^0$
and the nominal $B^0$ mass value is 0.7 MeV/$c^2$ \cite{PDG}.  This measurement
is the best
calibration available to establish the mass scale of the baryons
measured with hyperons in the final state,
because it involves a $J/\psi$ and displaced tracks.  Therefore, we 
use this $B^0$ mass discrepancy to establish the 
systematic uncertainty on the $\Xi_b^-$ and $\Omega_b^-$ mass measurements.
For the $B^0 \rightarrow J/\psi \, K_s^0$ mass measurement, 
approximately 3595 MeV/$c^2$ 
is taken up by the masses of the daughter particles.
The remaining 1685 MeV/$c^2$  is measured by the tracking system.  
This measured mass contribution is approximately 1370 MeV/$c^2$ for the
$\Xi_b^-$ and 1290 MeV/$c^2$ for the $\Omega_b^-$, corresponding to $\sim80\%$
of the $B^0$ value.  Consequently, we take this fraction of the $B^0$ 
mass measurement discrepancy to give an estimated systematic
uncertainty of 
0.55 MeV/$c^2$ for the $\Xi_b^-$ and $\Omega_b^-$ mass scale.

A shift of 0.5 MeV/$c^2$ is seen in our mass measurement of the
$\Lambda^0_b$, depending on whether the $\sigma_i^m$ used in the fit
is a constant 12 MeV/$c^2$ or is calculated for each event, 
based on the track parameter uncertainties.  This effect is not
statistically significant, but could appear in the 
$\Xi_b^-$ and $\Omega_b^-$ mass calculations. Therefore, it is considered to be 
a systematic uncertainty.  In addition, variations of $\pm0.3$ MeV/$c^2$
appear if the uncertainty scale factor $s_m$ is varied over  
the range $1.1-1.5$.  Finally, the $\Xi_b^-$ and $\Omega_b^-$ mass 
calculations depend on the rest masses of the decay daughters, 
since mass constraints are used in the candidate fit.  Only 
uncertainty on the mass
of the $\Omega^-$, which is known to $\pm0.3$ MeV/$c^2$ \cite{PDG},
contributes significantly.  
The quadrature sum of these effects is taken to obtain the final systematic
uncertainty of 0.8 MeV/$c^2$ for the $\Xi^-_b$ mass measurement, and 
0.9 MeV/$c^2$ for the $\Omega^-_b$ mass measurement.
The mass of the $\Xi^-_b$ is found to be 
$5790.9\pm2.6(\textrm{stat.})\pm0.8(\textrm{syst.})$ MeV/$c^2$, which is in
agreement with, and supersedes, our previous measurement \cite{CDF_Xi_b}.
The mass of the $\Omega^-_b$ is measured to be
$6054.4\pm6.8(\textrm{stat.})\pm0.9(\textrm{syst.})$ MeV/$c^2$. 
This value is consistent with
most predictions of the $\Omega^-_b$ mass, which fall in the range
$6010-6070$ GeV/$c^2$ \cite{Jenkins}.

\subsection{Lifetime Measurements \label{sect:life}}
The lifetime of $b$-hadrons is measured in this analysis by a technique
that is insensitive to the detailed lifetime characteristics of the
background.  This allows a lifetime calculation to be performed on a
relatively small sample, since a large number of events is not needed
for a background model to be developed.
The data are binned in $ct$, and the number of signal candidates in each
$ct$ bin is compared to the value that is expected for a particle with 
a given lifetime and measurement resolution.  

The calculation begins by expanding Eq.\ (\ref{eq:unbinned}) into
a  form that is binned in $ct$.  We maximize a 
likelihood function of the form
\begin{equation}
{\cal L} = \prod_{j=1}^{N_b}\prod_{i=1}^{N_j}  \left[ 
f_jG(m_i,m_0,s_m\sigma^m_i) + (1-f_j)  P_j^1(m_i)\right ],
\label{eq:binned}
\end{equation}
where $N_b$ is the number of $ct$ bins,
$N_j$ is the number of candidates in bin $j$, $f_j$ is the signal fraction 
found for bin $j$, and $P^1_j(m_i)$ is a first order polynomial for 
bin $j$ that describes the background.  This fit finds a single value of 
mass and resolution for all the data, and provides a best estimate
of the number of candidates in each $ct$ range.  

The maximization of Eq. (\ref{eq:binned}) provides a fraction $R_j$ of the
total signal in $ct$ bin $j$ given by $R_j = f_jN_j/\sum_{i=1}^{N_b}f_iN_i$
and its measurement uncertainty $\sigma_{R_j}$.
The lifetime $\tau$ can then be calculated by maximizing the likelihood
function given by
\begin{equation}
{\cal L} = \prod_{j=1}^{N_b} G(R_j,w_j,\sigma_{R_j})
\label{eq:life}
\end{equation}
where 
$w_j$ is the fraction of the signal that is calculated to occupy
bin $j$.  The measured lifetime distribution of $b$-hadrons
is a resolution-smeared exponential, given by Eq. (\ref{eq:smear}).
The expected content of each $ct$ bin is then given by
$w_j=\int^{ct^j_{high}}_{ct^j_{low}}
{\cal S}((ct),c\tau,\sigma^{(ct)}) d(ct) $ where $ct^j_{high}$
and $ct^j_{low}$ are the boundaries of $ct$ bin $j$.

In this application of the lifetime calculation, five bins in $ct$ were
used for all samples except the $\Omega^-_b$, where the
small sample size motivated the use of four bins.  
Studies with the $B^0$ sample indicate that little additional 
precision is gained by using more than five $ct$ bins.  The bin boundary 
between the lowest two bins was chosen to be 
$ct^1_{high}=100 \, \mu$m.  This choice has the
effect of placing the largest fraction of the combinatorial background 
into the first bin.
The remaining bin boundaries were chosen to place an equal number of 
candidates into each remaining bin, 
assuming they follow an exponential distribution
with a characteristic lifetime 
given by the initial value, $c\tau_{init}$, chosen for the fit.  
This algorithm gives the lower bin edges for the second and subsequent bins at 
$ct^j_{low} = ct^1_{high} - 
c\tau_{init} \ln \left (\frac{N_{b}-j}{N_{b}-1} \right )$. 
The lowest (highest) bin is unbounded on the low (high) side.

\begin{figure}[p]
\psfig{figure=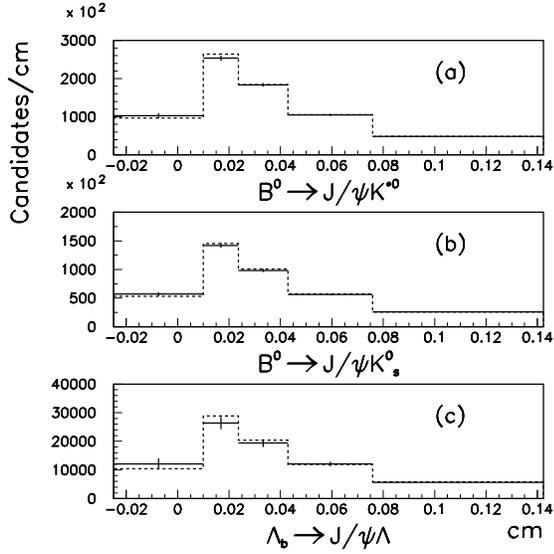, width=3.0in}
\caption{The solid histograms represent the number of 
(a) $B^0 \rightarrow J/\psi \, K^{*}(892)^{0}$,
(b) $B^0 \rightarrow J/\psi \, K^{0}_s$, and 
(c) $\Lambda^0_b \rightarrow J/\psi \, \Lambda$ 
candidates found in each $ct$ bin.  The dashed histogram is the
fit value.  Yields and fit values are normalized to candidates per cm,
and the bin edges are indicated.  The highest and lowest bins are not
bounded, but are truncated here for display purposes.
\label{fig:prd_b0_life}}
\end{figure}

\begin{figure}[p]
\psfig{figure=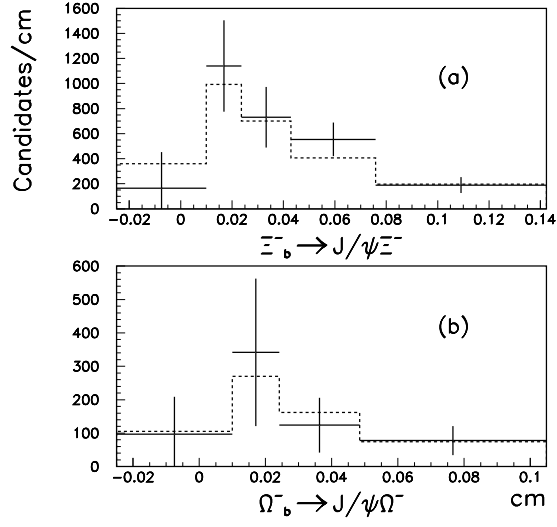, width=3.0in}
\caption{The solid histograms represent the number of 
(a) $\Xi^-_b \rightarrow J/\psi \, \Xi^-$ and
(b) $\Omega^-_b \rightarrow J/\psi \, \Omega^-$
candidates found in each $ct$ bin.  The dashed histogram is the
fit value.  Yields and fit values are normalized to candidates per cm,
and the bin edges are indicated.  The highest and lowest bins are not
bounded, but are truncated here for display purposes.
\label{fig:prd_xib_life} }
\end{figure}

All final states used in this analysis have three or more SVX\ II hits on 
each muon track, but not on any of the other tracks in the reconstruction.
This provides a comparable $ct$ resolution across the final states,
which falls in the range $15  \, \mu$m $< \sigma^{ct}_i <$ $40 \, \mu$m.
The average value of $\sigma^{ct}_i$ 
obtained from the $B^0$ and $\Lambda^0_b$ candidates is 30 $\mu$m, and this 
value was used in the lifetime fits. 
The signal yields and lifetimes obtained by maximizing
 Eq. (\ref{eq:life}) appear in Table\ \ref{table:life} along with the
statistical uncertainties on these quantities.
Comparisons between the number of candidates in each $ct$ bin
and the fit values are shown in Figs.\ \ref{fig:prd_b0_life} and
\ref{fig:prd_xib_life}.
The fits for the $B^0$ and $\Lambda^0_b$ were repeated for a variety of
different  $\sigma^{ct}_i$ over the range from 0 to $60 \, \mu$m.  
The resulting value of
$c\tau$ varied by $\pm2 \, \mu$m, which is taken as a 
systematic uncertainty
due to the treatment of  $\sigma^{ct}_i$.
The $B^0$ and $\Lambda^0_b$ $c\tau$ varied by $\pm5 \, \mu$m for 
different choices of
$N_b$, so this is considered an additional possible systematic
uncertainty.  
No systematic effect has been seen due to the choice of $c\tau_{init}$,
which was chosen to be 475 $\mu$m for the $B^0$, $\Lambda^0_b$ and $\Xi^-_b$, 
and 250 $\mu$m for the $\Omega^-_b$.
Systematic effects due to the detector mis-alignment are
estimated not to exceed $1 \, \mu$m.  The estimates of these effects, 
combined in quadrature,
provide a systematic uncertainty of $6 \, \mu$m on the $B^0$ lifetime
measurements, a relative uncertainty of 1.3\%.
The results of the $B^0$ lifetime measurements are
consistent with the nominal value of $459 \, \pm \, 6 \, \mu$m \cite{PDG},
which serves as a check on the analysis technique.  In addition, the
lifetime result obtained here for the $\Lambda^0_b$ is consistent with our 
previous measurement \cite{CDF_LbLife}, 
which was based on a continuous lifetime fit similar to 
Eq. (\ref{eq:unbinned_life}).
\begin{table}[ht]
\begin{center}
\caption{Signal yields and lifetimes obtained for the $b$-hadrons.
\label{table:life}}
\begin{tabular}{ccc}
\hline \hline
Resonance  & Yield &$c\tau$ ($\mu$m) \\
\hline
 $B^0 (J/\psi \, K^{*}(892)^{0})$ &  $17250\pm305$ &  $453\pm6$  \\
 $B^0 (J/\psi \, K^0_s) $ &   $9424\pm167$ &  $448\pm7$  \\
 $\Lambda^0_b$              &   $1934\pm93$  &  $472\pm17$ \\
 $\Xi^-_b $               &     $66^{+14}_{-9}$  &  $468^{+82}_{-74}$ \\
 $\Omega^-_b$             &     $16^{+6}_{-4}$   &  $340^{+160}_{-120}$ \\
\hline \hline
\end{tabular}
\end{center}
\end{table}
Consequently, a systematic uncertainty of 1.3\% of the central 
lifetime value
is taken for  the  $b$-baryon lifetime measurements.
We measure the lifetime of the $\Xi^-_b$ to be
$1.56^{+0.27}_{-0.25}(\textrm{stat.})\pm0.02(\textrm{syst.})$ ps 
and the lifetime of the $\Omega^-_b$ to be
$1.13^{+0.53}_{-0.40}(\textrm{stat.})\pm0.02(\textrm{syst.})$ ps.

\subsection{Relative Production Rate Measurements}

A further goal of this analysis
is to measure the production rates of the $\Xi_b^-$ and $\Omega_b^-$,
relative to the more plentiful $\Lambda^0_b$, where
we measure ratios of 
cross section times branching fractions.  In the case of the $\Xi^-_b$,
we evaluate
\begin{eqnarray}
\lefteqn{\frac{\sigma(\Xi_b^-){\cal B}(\Xi_b^- \rightarrow J/\psi \, \Xi^-)
 {\cal B}(\Xi^- \rightarrow \Lambda \, \pi^-)}
     {\sigma(\Lambda^0_b){\cal B}(\Lambda^0_b \rightarrow J/\psi \, \Lambda)} =}
\nonumber \\ 
& & \frac{N_{data}(\Xi_b^- \rightarrow J/\psi \, \Xi^-)}
     {N_{data}(\Lambda^0_b \rightarrow J/\psi \,\Lambda )}
\frac{\epsilon_{\Lambda^0_b}}{\epsilon_{\Xi_b^-}}.
\label{eq:rel_rate}
\end{eqnarray}
where $\sigma(h)$ is the production cross section of hadron $h$, ${\cal B}$
corresponds to the indicated branching fractions, $N_{data}$
are the number of indicated candidates seen in the data, and
$\epsilon_h$ is the acceptance and reconstruction efficiency for hadron $h$.
A similar expression for the $\Omega^-_b$ applies as well.

The hyperon branching fractions are well measured, and we use the nominal 
values for these quantities \cite{PDG}.  The number of events for each 
state is obtained from the lifetime fit technique described previously
(Sec.\ \ref{sect:life}) and listed in Table\ \ref{table:life}.  
The acceptance and efficiency terms require careful consideration
because the acceptance of the CDF tracking system is not well modeled
for tracks with $p_T \, < \, 400$ MeV/$c$.  Consequently,
the calculation of total acceptance is dependent on the assumed $p_T$
distribution of the particle of interest.  Simple application of
our simulation to estimate the total efficiency would leave the
results with a dependence on the underlying generation model \cite{NDE}
which is difficult to estimate.
Therefore,
a strategy has been adopted to reduce the sensitivity of the relative rate
measurement to the simulation assumptions.  This method divides the
data into subsets, defined by limited ranges of $p_T$.  The efficiency over 
a limited range of $p_T$ can be calculated more reliably, since the 
variation of a reasonable simulation model, such as the one used here, is small
over the limited $p_T$ range.

As was done with the mass and lifetime measurements, the $B^0$ sample is
used as a reference point for the relative rate measurement.  In analogy to 
Eq.\ (\ref{eq:rel_rate}), the ratio of branching fractions for the $B^0$
is given by
\begin{eqnarray}
\frac{{\cal B}(B^0 \rightarrow J/\psi \, K^0)
      {\cal B}(K^0 \rightarrow K^0_s)
      {\cal B}(K^0_s \rightarrow \pi^+ \, \pi^-)}
{{\cal B}(B^0 \rightarrow J/\psi \, K^{*}(892)^{0})
 {\cal B}(K^{*}(892)^{0} \rightarrow K^+ \, \pi^-)} = \nonumber \\ 
\frac{N_{data}(B^0 \rightarrow J/\psi \, K^0_s)}
     {N_{data}(B^0 \rightarrow J/\psi \, K^{*}(892)^{0})}
\frac{\epsilon_{K^{*}(892)^{0}}}{\epsilon_{K^0_s}}.
\end{eqnarray}
The branching fractions are taken to be 
${\cal B}(K^0 \rightarrow K^0_s)=0.5$, 
${\cal B}(K^{*}(892)^{0} \rightarrow K^+ \, \pi^-)=2/3$ 
and ${\cal B}(K^0_s \rightarrow \pi^+ \, \pi^-)=0.692$ \cite{PDG}.  
The number of 
candidates for each final state obtained for several $p_T$ ranges
is then combined with the acceptance and reconstruction efficiency for that
range to obtain the ratio of branching fractions indicated in 
Table\ \ref{table:bzero_result}.  The full range  of $6 - 20$ GeV/$c$
was chosen to correspond to the range of data available in
the $\Xi^-_b$ and $\Omega^-_b$ samples. 
These results are consistent with the
nominal value of $0.655\pm0.038$ \cite{PDG}
for the branching fraction ratio, and provide confirmation of the
accuracy of the detector simulation for these states.
\begin{table*}[hbt]
\begin{center}
\caption{The yields of $B^0$ candidates obtained for several
ranges of $p_T(B^0)$ and the branching fraction ratio
obtained for each subset.  \protect\label{table:bzero_result}}
\vspace{3mm}
\begin{tabular}{cccc}
\hline \hline
$p_T$(GeV/$c$) & $B^0 \rightarrow J/\psi K^{*}(892)^{0}$ &
$B^0 \rightarrow J/\psi K^0_s$ & 
$\frac{{\cal B}(B^0 \rightarrow J/\psi \, K^0)}
      {{\cal B}(B^0 \rightarrow J/\psi \, K^{*}(892)^{0})}$ \\
\hline
$6-7.5$   & $2640\pm74$ & $1196\pm23$ & $0.59\pm0.04$ \\
$7.5-9$   & $2687\pm52$ & $1361\pm50$ & $0.64\pm0.03$ \\
$9-11$    & $3189\pm49$ & $1685\pm34$ & $0.63\pm0.03$ \\
$11-14$   & $3243\pm54$ & $1615\pm50$ & $0.64\pm0.03$ \\
$14-20$   & $2787\pm56$ & $1321\pm27$ & $0.63\pm0.03$ \\
\hline
$6-20$  & $14546\pm129$ & $7178\pm98$ & $0.628\pm0.014$ \\
\hline \hline
\end{tabular}
\end{center}
\end{table*}

The samples of $\Xi^-_b$ and $\Omega^-_b$ are too small to be divided
into ranges of $p_T$, as is done for the $B^0$.  Therefore, the 
acceptance and reconstruction efficiency must be obtained over the
wider range of 6-20 GeV/$c$, and a production distribution 
as a function of $p_T$ must be assumed over this range.
The production distribution used here is derived from the data, rather
than adopting a theoretically motivated model.
The derivation assumes that the  $\Xi^-_b$ and $\Omega^-_b$ 
are produced with the same $p_T$ distribution as the
$\Lambda^0_b$.  We then use the observed  $p_T$  distribution of 
$\Lambda^0_b$ production
to obtain the total efficiency for the $\Xi^-_b$ and $\Omega^-_b$ states.

The first step in obtaining the total acceptance and reconstruction 
efficiency terms is to divide the $\Lambda^0_b$ sample into several ranges
of $p_T$.  The number of candidates is found by fitting each sample
with the likelihood defined in Eq. (\ref{eq:unbinned}).  The reconstruction
efficiency for the $\Lambda^0_b$ in each range of $p_T$ was obtained 
by simulating events through the full detector simulation.  The yield 
and efficiency are then combined to give a quantity that is proportional
to $\sigma(\Lambda^0_b) {\cal B}(\Lambda^0_b \rightarrow J/\psi \, \Lambda)$
for each range of $p_T$.  
The acceptance and reconstruction efficiency terms 
for each $p_T$ range, $\epsilon_{\Xi_b}(p_T)_j$, are simply
obtained from the simulation.  
The total reconstruction 
efficiency over the full range of $p_T$ is 
$\epsilon_{\Xi_b} = \sum_j^{N_j} f_j^{\Lambda^0_b}\epsilon_{\Xi_b}(p_T)_j$
where 
$N_j$ is the number of $p_T$ ranges, and
$f_j^{\Lambda^0_b}$ is the fraction of the $\Lambda^0_b$ produced
in $p_T$ range $j$.  
These factors and their statistical uncertainties
appear in Table\ \ref{table:xib_effic}.  
\begin{table*}[hbt]
\begin{center}
\caption{The efficiencies
of $\Xi_b$ and $\Omega_b$ candidates obtained for several
ranges of $p_T$ and the fraction of $\Lambda^0_b$
events produced for each range.  
For the total efficiency over the $p_T$ range $6-20$ GeV/$c^2$,
the first uncertainty term is due to the $\Lambda_b^0$ sample, 
and the second is due to the simulation sample size.
\protect\label{table:xib_effic}}
\vspace{3mm}
\begin{tabular}{ccccc}
\hline \hline
$p_T$(GeV/$c$) & $f_j^{\Lambda^0_b}$ & 
$\epsilon_{\Lambda^0_b}(p_T) \times 10^{-2}$ &
$\epsilon_{\Xi_b}(p_T) \times 10^{-3}$ &
$\epsilon_{\Omega_b}(p_T) \times 10^{-3}$\\
\hline
$6-7.5$ & $0.411\pm0.031$ & $1.40\pm0.04$ & $2.37\pm0.14$  & $2.21\pm0.17$ \\
$7.5-9$   & $0.277\pm0.020$ & $2.59\pm0.06$ & $4.96\pm0.28$  & $6.73\pm0.41$ \\
$9-11$    & $0.168\pm0.011$ & $4.14\pm0.10$ & $9.40\pm0.44$  & $11.54\pm0.61$ \\
$11-14$   & $0.092\pm0.006$ & $6.39\pm0.14$ & $16.08\pm0.71$  & $23.26\pm1.02$ \\
$14-20$   & $0.052\pm0.005$ & $9.32\pm0.22$ & $24.19\pm1.11$ & $40.27\pm1.96$ \\
\hline
$6-20$  & & $3.07\pm0.14\pm0.04$ & $6.67\pm0.22\pm0.17$ & 
$8.96\pm0.32\pm0.24$ \\
\hline \hline
\end{tabular}
\end{center}
\end{table*}
The $p_T$ integrated acceptance and efficiency terms 
are then used to solve
Eq. (\ref{eq:rel_rate}) for the relative rates of production.  
The $\Lambda^0_b$ yield in the $p_T$ range of $6-20$ GeV/$c$ is
$1812\pm61$,
while $66^{+16}_{-9}$ and $16^{+6}_{-4}$
are found for the  $\Xi_b^-$ and $\Omega_b^-$,
respectively.
The relative production ratios
are $0.167^{+0.037}_{-0.025}$ for the $\Xi_b^-$ and
$0.045^{+0.017}_{-0.012}$ for the $\Omega_b^-$ where these uncertainties are 
statistical, and contain the contributions from the 
$\Lambda^0_b$ measurements.
  
The total uncertainty on the efficiency contains contributions from both
the calculation of $f_j^{\Lambda^0_b}$ and the size of the sample used for the 
simulation.  These contributions
were added, to obtain a total relative uncertainty on
the efficiency terms of 6\%. 
The simulation of the tracking system is accurate to within 3\% for the 
five-track final states used in this analysis \cite{CDF_Bcross}. An
additional 0.3\% is assigned to the $\Omega^-$, due to
our characterization of the material in the detector
and its effect on the $K^-$ tracking efficiency.  
The uncertainty on the $\Xi^-$ 
branching fraction does not contribute significantly,
and the $\Omega^-$ branching fraction is known to within 1\%.
The mass of the $\Omega_b^-$ used in the simulation was varied 
over the range $6.0-6.19$ GeV/$c^2$, and the efficiency
calculations were repeated. The efficiency was found to 
remain constant to within 5\%.  We assign this value as an additional
systematic uncertainty on the $\Omega_b^-$ efficiency.
An additional systematic uncertainty of 2.5\% 
due to the $\Lambda_b^0$ yield is obtained
by varying $c\tau(\Lambda^0_b)$ over a $\pm50$ $\mu$m range.
These systematic effects were combined in quadrature to provide an estimate
for the total relative systematic uncertainty on the production ratios
of 7\% for the $\Xi^-_b$ and 9\% for the  $\Omega_b^-$.
Our measurements of the relative production rates are
$\frac{\sigma(\Xi_b^-){\cal B}(\Xi_b^- \rightarrow J/\psi \, \Xi^-)}
     {\sigma(\Lambda^0_b){\cal B}(\Lambda^0_b \rightarrow J/\psi \, \Lambda)} =
0.167^{+0.037}_{-0.025}(\textrm{stat.})\pm0.012(\textrm{syst.})$
and 
$\frac{\sigma(\Omega_b^-){\cal B}(\Omega_b^- \rightarrow J/\psi \, \Omega^-)}
      {\sigma(\Lambda^0_b){\cal B}(\Lambda^0_b \rightarrow J/\psi \, \Lambda)} =
0.045^{+0.017}_{-0.012}(\textrm{stat.})\pm0.004(\textrm{syst.})$
for the $\Xi^-_b$ and   $\Omega_b^-$, respectively.

\section{Conclusions \label{sect:Conclusions}}
In conclusion, we have used data collected with 
the CDF II detector at the Tevatron to
observe the $\Omega^-_b$ in $p\overline{p}$ collisions.
The reconstruction used for this observation
and the techniques for measuring the properties of the $\Omega^-_b$
are used on other $b$-hadron properties that have been measured
previously, which provide a precise calibration for the analysis.
A signal of $16^{+6}_{-4}$ $\Omega^-_b$ candidates,  
with a significance equivalent to $5.5\sigma$ when combining both
mass and lifetime information, is seen in the decay channel
$\Omega^-_b \rightarrow J/\psi \, \Omega^-$ with
$J/\psi \rightarrow \mu^+ \, \mu^-$,
$\Omega^- \rightarrow \Lambda \, \pi^-$, and
$\Lambda \rightarrow p \, \pi^-$.  The mass of this baryon
is measured to be $6054.4\pm6.8 (\textrm{stat.}) \pm0.9 (\textrm{syst.})$
MeV/$c^2$,
which is consistent with theoretical expectations 
\cite{Jenkins}.
In addition, we measure the lifetime of the $\Omega^-_b$ to be
$1.13^{+0.53}_{-0.40} (\textrm{stat.}) \pm0.02 (\textrm{syst.})$ ps, and 
the $\Omega^-_b$ production
relative to the $\Lambda^0_b$ to be
$\frac{\sigma(\Omega_b^-){\cal B}(\Omega_b^- \rightarrow J/\psi \, \Omega^-)}
      {\sigma(\Lambda^0_b){\cal B}(\Lambda^0_b \rightarrow J/\psi \, \Lambda)} =
0.045^{+0.017}_{-0.012}(\textrm{stat.})\pm0.004(\textrm{syst.})$.
The additional data available to this analysis allows an update to
our previous $\Xi^-_b$ mass measurement \cite{CDF_Xi_b}.
A new value of
$5790.9\pm2.6  (\textrm{stat.}) \pm0.8 (\textrm{syst.})$ MeV/$c^2$
is obtained for the $\Xi^-_b$ mass.
The lifetime of the $\Xi^-_b$ is measured to be
$1.56^{+0.27}_{-0.25} (\textrm{stat.}) \pm0.02 (\textrm{syst.})$  ps, 
which is the first measurement
of this quantity in a fully reconstructed final state.  
Finally, the relative production of the $\Xi^-_b$
compared to the $\Lambda^0_b$ is found to be
$\frac{\sigma(\Xi_b^-){\cal B}(\Xi_b^- \rightarrow J/\psi \, \Xi^-)}
     {\sigma(\Lambda^0_b){\cal B}(\Lambda^0_b \rightarrow J/\psi \, \Lambda)} =
0.167^{+0.037}_{-0.025}(\textrm{stat.})\pm0.012(\textrm{syst.})$.

The first reported observation of the $\Omega^-_b$ measured a mass
of $6165\pm10(\textrm{stat.})\pm13(\textrm{syst.})$  MeV/$c^2$ 
\cite{D0_Omega_b}.
The mass measurement presented here differs from Ref. \cite{D0_Omega_b} by
$111\pm12(\textrm{stat.})\pm14(\textrm{syst.})$ MeV/$c^2$, 
where we have combined the 
statistical uncertainties of the two measurements in quadrature,
and summed the systematic uncertainties.  The two measurements
appear to be inconsistent.

The relative rate measurement 
presented in Ref. \cite{D0_Omega_b} is
$\frac{f(b \rightarrow \Omega_b^-)
    {\cal B}(\Omega_b^- \rightarrow J/\psi \, \Omega^-)}
     {f(b \rightarrow \Xi_b^-)
    {\cal B}(\Xi_b^- \rightarrow J/\psi \, \Xi^-)} =
0.80\pm0.32(\textrm{stat.})^{+0.14}_{-0.22}(\textrm{syst.})$ where 
$f(b \rightarrow \Omega_b^-)$ and $f(b \rightarrow \Xi_b^-)$ are the 
fractions of $b$ quarks that hadronize to $\Omega^-_b$ and $\Xi^-_b$.
The equivalent quantity taken from the present analysis is
$\frac{\sigma(\Omega_b^-){\cal B}(\Omega_b^- \rightarrow J/\psi \, \Omega^-)}
      {\sigma(\Xi_b^-){\cal B}(\Xi_b^- \rightarrow J/\psi \, \Xi^-)} =
0.27\pm0.12(\textrm{stat.})\pm0.01(\textrm{syst.})$.  
Neither measurement is very precise, since 
a ratio is taken of two small samples.  Nevertheless,
this analysis indicates a rate of $\Omega_b^-$ production substantially lower
than Ref. \cite{D0_Omega_b}.
Consequently,
the analysis presented here is not able to confirm the 
$\Omega^-_b$ observation
reported in Ref. \cite{D0_Omega_b}.  Future work 
is needed to resolve the discrepancy between the two results.

We thank the Fermilab staff and the technical staffs of the participating
 institutions for their vital contributions. This work was supported by 
the U.S. Department of Energy and National Science Foundation; the Italian
 Istituto Nazionale di Fisica Nucleare; the Ministry of Education, Culture,
 Sports, Science and Technology of Japan; the Natural Sciences and Engineering
 Research Council of Canada; the National Science Council of the Republic 
of China; the Swiss National Science Foundation; the A.P. Sloan Foundation;
 the Bundesministerium f\"ur Bildung und Forschung, Germany; the Korean 
Science and Engineering Foundation and the Korean Research Foundation; 
the Science and Technology Facilities Council and the Royal Society, UK; 
the Institut National de Physique Nucleaire et Physique des Particules/CNRS; 
the Russian Foundation for Basic Research; the Ministerio de Ciencia e
 Innovaci\'{o}n, and Programa Consolider-Ingenio 2010, Spain; 
the Slovak R\&D Agency; and the Academy of Finland.

\end{document}